\documentclass[apj]{emulateapj}

\usepackage{apjfonts}
\usepackage{lscape}

\def\ltsima{$\; \buildrel < \over \sim \;$}
\def\simlt{\lower.5ex\hbox{\ltsima}}
\def\gtsima{$\; \buildrel > \over \sim \;$}
\def\simgt{\lower.5ex\hbox{\gtsima}}

\newcommand{\suzaku}{{\it Suzaku}}
\newcommand{\swone}{{Swift J0138.6--4001}}
\newcommand{\swtwo}{{Swift J0255.2--0011}}
\newcommand{\swthree}{{Swift J0350.1--5019}}
\newcommand{\swfive}{{Swift J0505.7--2348}}
\newcommand{\swsix}{{Swift J0601.9--8636}}
\newcommand{\swsixteen}{{Swift J1628.1--5145}}

\newcommand{\two}{{J0255.2--0011}}
\newcommand{\three}{{J0350.1--5019}}
\newcommand{\five}{{J0505.7--2348}}
\newcommand{\six}{{J0601.9--8636}}
\newcommand{\sixteen}{{J1628.1--5145}}

\newcommand{\idone}{{ESO 297--G018}}
\newcommand{\idtwo}{{NGC 1142}}
\newcommand{\idthree}{{2MASX J03502377--5018354}}
\newcommand{\idfive}{{2MASX J05054575--2351139}}
\newcommand{\idsix}{{ESO 005--G004}}
\newcommand{\idsixteen}{{Mrk 1498}}

\newcommand{\swift}{{\it Swift}}
\newcommand{\integral}{{\it INTEGRAL}}
\newcommand{\xmm}{{\it XMM-Newton}}

\newcommand{\nh}{$N_{\rm H}$}
\newcommand{\cosmo}{($H_0$, $\Omega_{\rm m}$, $\Omega_{\lambda}$)}

\newcommand{\Swift}{{\it Swift}}
\newcommand{\Suzaku}{{\it Suzaku}}

\shorttitle{{\Suzaku} View of {\Swift}/BAT AGNs (I)}
\shortauthors{Eguchi et al.}

\begin{document}

\title{{\Suzaku} View of the {\Swift}/BAT Active Galactic Nuclei (I):\\
Spectral Analysis of Six AGNs\\ and Evidence for Two Types of Obscured Population
}
\author{
 Satoshi Eguchi\altaffilmark{1},
 Yoshihiro Ueda\altaffilmark{1},
 Yuichi Terashima\altaffilmark{2},
 Richard Mushotzky\altaffilmark{3},
 Jack Tueller\altaffilmark{3}
}

\altaffiltext{1}{Department of Astronomy, Kyoto University, Kyoto 606-8502, Japan}
\altaffiltext{2}{Department of Physics, Faculty of Science, Ehime University, Matsuyama 790-8577, Japan}
\altaffiltext{3}{NASA/Goddard Space Flight Center, Greenbelt, MD 20771, USA}

\begin{abstract} 

We present a systematic spectral analysis with \suzaku\ of six AGNs
detected in the \swift /BAT hard X-ray (15--200 keV) survey, \swone,
\two, \three, \five, \six, and \sixteen. This is considered to be a
representative sample of new AGNs without X-ray spectral information
before the BAT survey. We find that the 0.5--200 keV spectra of these
sources can be uniformly fit with a base model consisting of heavily
absorbed (log $N_{\rm H} >$23.5 cm$^{-2}$) transmitted components,
scattered lights, a reflection component, and an iron-K emission
line. There are two distinct groups, three ``new type'' AGNs
(including the two sources reported by \citealt{Ueda2007}) with an
extremely small scattered fraction ($f_{\rm scat}<0.5\%$) and strong
reflection component ($R=\Omega/2\pi \gtrsim 0.8$ where $\Omega$ is the
solid angle of the reflector), and three ``classical type'' ones with
$f_{\rm scat}>0.5\%$ and $R \lesssim 0.8$. The spectral parameters
suggest that the new type has an optically thick torus for Thomson
scattering ($N_{\rm H} \sim 10^{25}$ cm$^{-2}$) with a small opening
angle $\theta\sim 20^\circ$ viewed in a rather face-on geometry, while
the classical type has a thin torus ($N_{\rm H} \sim 10^{23-24}$
cm$^{-2}$) with $\theta\gtrsim 30^\circ$. We infer that a significant
number of new type AGNs with an edge-on view is missing in the current
all-sky hard X-ray surveys.

\end{abstract}


\section{Introduction}

The growth history of supermassive black holes (SMBHs) in galactic
centers is a key question for understanding cosmic history. This can
be traced by observations of Active Galactic Nuclei (AGNs), where mass
accretion onto SMBHs is taking place. Many observations indicate that
obscured AGNs are the major AGN population both in the local universe
and at intermediate-to-high redshifts \citep[see
e.g.,][]{Hasinger2008}.  In fact, much of the SMBH growth is
theoretically expected to happen during an obscured phase in the
evolution of AGNs \citep[e.g.,][]{Hopkins2006}. Thus, detecting all
the obscured AGNs in the universe and revealing their nature is a
fundamental issue to be addressed by modern astrophysics. Our
knowledge of this population is still quite limited even in the local
universe, due the difficulty of detecting highly absorbed objects in
most wavelength bands.

\begin{deluxetable*}{cccccc}
\tablenum{1}
\tablecaption{List of Targets\label{tab-targets}}
\tablewidth{0pt}
\tablehead{\colhead{SWIFT} & \colhead{Optical/IR Identification}
 & \colhead{R.A. (J2000)} & \colhead{Dec. (J2000)}
 & \colhead{Redshift} & \colhead{Classification}}
\startdata
 J0138.6--4001 & ESO 297--G018 & 01 38 37.16 & -40 00 41.1 & 0.0252 & Seyfert 2 \\
 J0255.2--0011 & NGC 1142 & 02 55 12.196 & -00 11 0.81 & 0.0288 & Seyfert 2\\
 J0350.1--5019 & 2MASX J03502377--5018354 & 03 50 23.77 & -50 18 35.7 & 0.036 & Galaxy \\
 J0505.7--2348 & 2MASX J05054575--2351139 & 05 05 45.73 & -23 51 14.0 & 0.0350 & Seyfert 2 \\
 J0601.9--8636 & ESO 005--G004 & 06 05 41.63 & -86 37 54.7 & 0.0062 & Galaxy \\
 J1628.1$+$5145 & Mrk 1498 & 16 28 4.06 & +51 46 31.4 & 0.0547 & Seyfert 1.9 \\
\enddata
\tablecomments{The position, redshift, and classification for each source is taken from the NASA/IPAC Extragalactic Database.}
\end{deluxetable*} 

Obscured AGNs, often referred as type 2 AGNs, are characterized by
heavily absorbed X-ray spectra and/or by the absence of broad emission
lines in the optical spectra. The unified model of AGNs attributes
these differences to the blockage of the line of sight by a dusty
torus.  Recent X-ray and IR observations have revealed populations of
AGNs without any signature of AGNs in the optical band
\citep[e.g.,][]{Maiolino2003}.  This fact implies that AGN surveys
relying on optical emission signatures (such as the [O~III] $\lambda
5007$ line) may be incomplete.

Sensitive hard X-ray observations above 10 keV provide us with the
least biased AGN samples in the local universe, due to the strong
penetrating power against photo-electric absorption, except for
heavily Compton thick (column density of log \nh\ $\gtrsim$ 24.5
cm$^{-2}$) objects. Recent all sky hard X-ray surveys performed with
\swift /BAT \citep{Tueller2008} and \integral\ \citep{Bassani2006,
Krivonos2007} have started to detect many local AGNs with about 10
times better sensitivity than that of previous missions in this energy
band.  They have shown that roughly half of all local AGN are indeed
obscured AGNs. These hard X-ray samples contain a number of new AGNs,
for which detailed follow-up studies have not been done yet. Some of
these sources objects were not recognized to be AGNs previously even
though there were optical imaging and spectroscopy data available.

Our team has started to make follow-up observations of selected \swift
/BAT AGNs with the \suzaku\ satellite to obtain the best quality broad
band high energy data. The major purposes are (1) to discover new
populations of AGNs, (2) to measure their spectral properties, which
make it possible to construct the \nh\ distribution of a complete
\swift /BAT sample, and (3) to accurately constrain the strength of
the reflection component, which is an important yet uncertain
parameter in population synthesis models of the X-ray background in
explaining its broad-band shape \citep{Ueda2003, Gilli2007}. In
particular, the simultaneous coverage over the 0.2--70 keV band with
\suzaku\ is crucial to discriminate degeneracy of spectral models
under possible time variability and to determine the amounts of
scattered and reflected components, key information to constrain the
geometry of surrounding matter around the nucleus.

The first results obtained from the \suzaku\ follow-up of two \swift
/BAT AGNs led to the discovery of a new type of deeply buried AGNs
\citep{Ueda2007}. Here we present the results of a detailed analysis
of six \swift /BAT AGNs, including the two sources (\swone\ and \swsix
) already reported there\footnote{For these two sources, the results
of the present paper, which are based on more recent calibration of
the instruments and background estimation, supersedes those by
\citet{Ueda2007}, although the essence is not changed.}. This paper
constitutes the first one in the series of papers to be published on
\suzaku-\swift\ AGNs, followed by the second paper by Winter et al.\
(in preparation). \S~2 summarizes the sample and observations. In
\S~3, we mainly present the results of detailed spectral analysis with
\suzaku. Discussion is given in \S~4. We adopt the cosmological
parameters \cosmo\ = (70 km s$^{-1}$ Mpc$^{-1}$, 0.3, 0.7) throughout
the paper.

\section{Observations}

\begin{deluxetable*}{cccccc}
\tablenum{2}
\tablecaption{Observation Log\label{tab-observations}}
\tablewidth{0pt}
\tablehead{\colhead{SWIFT} & \colhead{Start Time (UT)} & \colhead{End Time}
 & \colhead{Exposure\tablenotemark{a} (XIS)} & \colhead{Exposure (HXD/PIN)} & \colhead{SCI\tablenotemark{b}}}
\startdata
J0138.6--4001 & 2006 Jun 04 18:13 & Jun 05 05:00 & 21.2 ks & 16.5 ks & Off \\
J0255.2--0011 & 2007 Jan 23 14:54 & Jan 26 05:30 & 101.6 ks & 80.6 ks & On \\
J0350.1--5019 & 2006 Nov 23 02:07 & Nov 23 13:33 & 19.3 ks & 15.7 ks & On \\
J0505.7--2348 & 2006 Apr 01 22:12 & Apr 04 02:55 & 78.6 ks & 50.3 ks & Off \\
J0601.9--8636 & 2006 Apr 13 16:24 & Apr 14 01:52 & 19.8 ks & 15.7 ks & Off \\
J1628.1$+$5145 & 2006 Apr 15 18:49 & Apr 16 09:22 & 23.6 ks & 19.5 ks & Off \\
\enddata
\tablenotetext{a}{Based on the good time interval for XIS-0.}
\tablenotetext{b}{With/without the spaced-row charge injection for the XIS\citep{Nakajima2008}.}
\end{deluxetable*}

\begin{figure*}
\epsscale{0.8}
\plotone{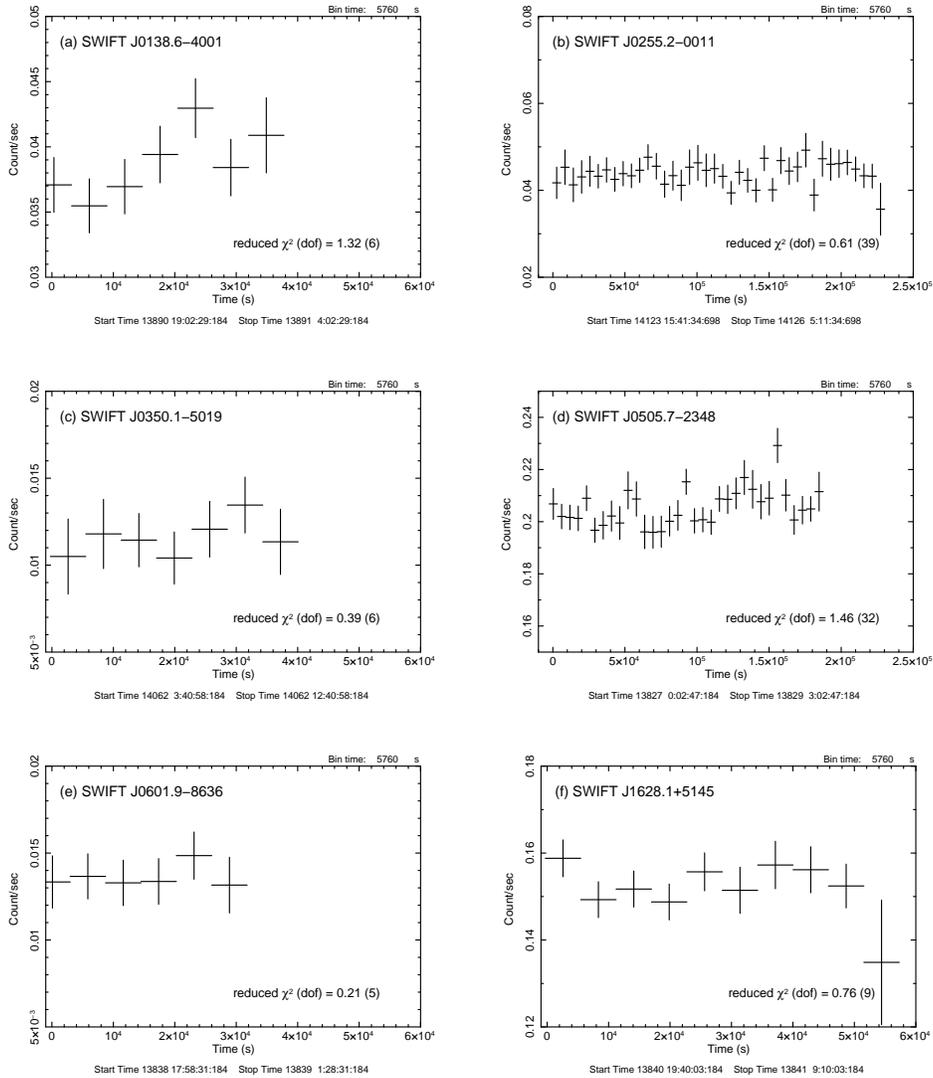}
\caption{The background subtracted light curves of XIS in the 2--10 keV band
during the \suzaku\ observations. One bin corresponds
to 96 minutes. The data from the XIS-0 and XIS-2 are summed for
\swtwo\ and \swthree , while those from XIS-0, XIS-2, and XIS-3 are
summed for the rest. The numbers listed in each panel represent the
value of reduced $\chi^2$ with the degree of freedom for the constant
flux hypothesis. 
\label{fig-lc_xis}} 
\end{figure*}

\begin{deluxetable*}{ccccccc}
\tablenum{3}
\tablecaption{Cutoff Energies ($E_{\rm{cut}}$) determined by the BAT spectra\label{tab-ecut}}
\tablewidth{0pt}
\tablehead{\colhead{$\Omega / 2 \pi$} & \colhead{SWIFT J0138.6--4001} & \colhead{J0255.2--0011} & \colhead{J0350.1--5019}
 & \colhead{J0505.7--2348} & \colhead{J0601.9--8636} & \colhead{J1628.1$+$5145}}
\startdata
$0$ & $> 317$ & $> 289$ & $> 106$ & $> 318$ & $> 116$ & $> 99$ \\
$\chi^{2} / \rm{d.o.f.}$ & $1.58 / 2$ & $4.57 / 6$ & $1.16 / 6$ & $13.58 / 6$ & $4.96 / 2$ & $2.38 / 6$ \\ \hline
$2$ & $329$ ($> 91$) & $> 253$ & $> 95$ & $> 306$ & $> 82$ & $> 87$ \\
$\chi^{2} / \rm{d.o.f.}$ & $1.05 / 2$ & $5.21 / 6$ & $2.40 / 6$ & $15.24 / 6$ & $3.37 / 2$ & $4.61 / 6$ \\
\enddata
\tablecomments{The unit of $E_{\rm{cut}}$ is keV.}
\end{deluxetable*}

\begin{figure*}
\epsscale{0.80}
\plotone{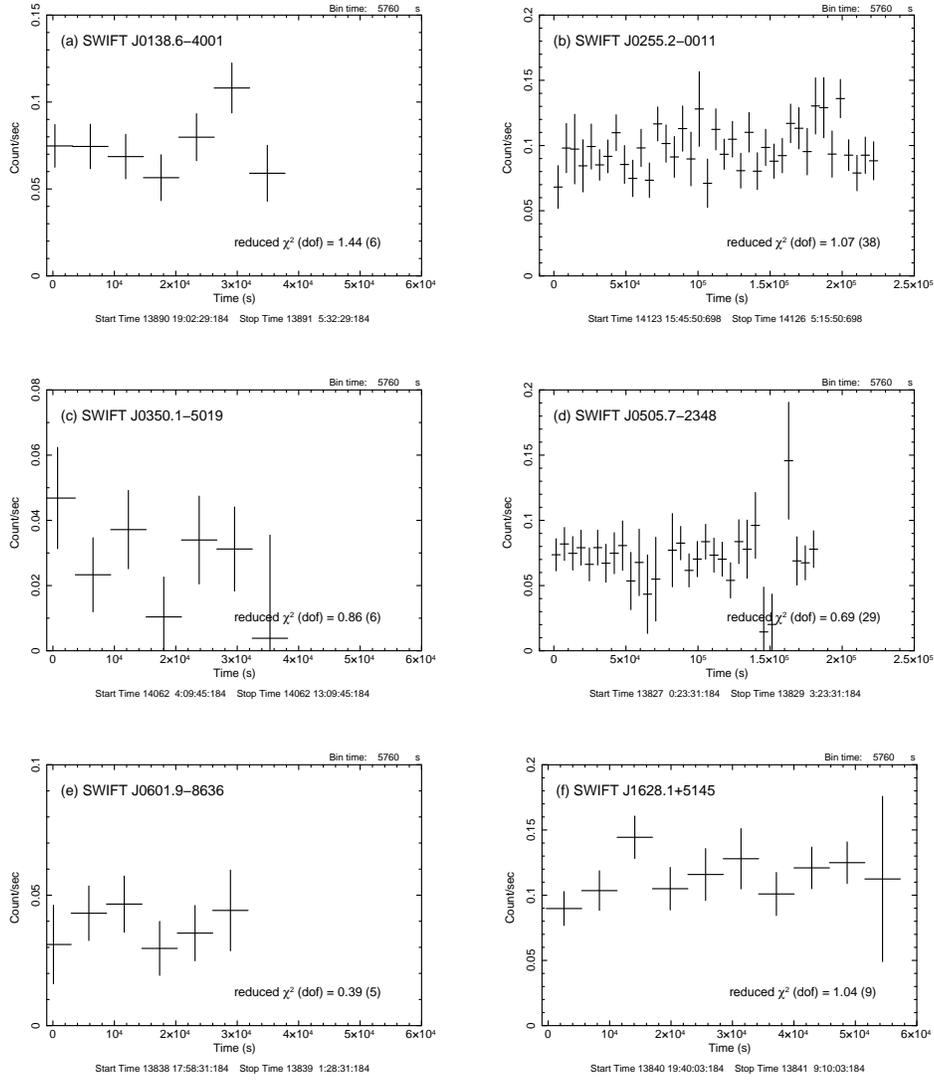}
\caption{The background subtracted light curves of HXD/PIN in the
15--40 keV band during the \suzaku\ observations.
One bin corresponds to 96 minutes.  The numbers listed in each panel
represent the value of reduced $\chi^2$ with the degree of freedom for
the constant flux hypothesis.  \label{fig-lc_pin}} \end{figure*}

We observed six \swift/BAT AGNs with {\Suzaku} between 2006 April and
2007 January in the AO-1 phase. The targets are \swone, \two, \three,
\five, \six, and \sixteen, whose identification in the optical or
infrared band is \idone, \idtwo, \idthree, \idfive, \idsix, and
\idsixteen, respectively \citep{Tueller2008}. The basic information of
our targets is summarized in Table~\ref{tab-targets}. 2MASX
J03502377--5018354 and ESO 005--G004 had not been identified as active
galaxies in the optical band, and revealed to contain an AGN for the
first time with the detection of hard X-rays by the BAT. The sources
were selected as the \suzaku\ targets because at the time of our
proposals these ``new'' \swift\ AGNs had no spectral information
available below 10 keV. Recently short follow-up observations of these
targets have been carried out with \xmm\ or the X-Ray Telescope (XRT)
onboard \swift\ \citep{Winter2008a}.

{\Suzaku}, the fifth Japanese X-ray satellite \citep{Mitsuda2007},
carries four X-ray CCD cameras called the X-ray Imaging Spectrometer
(XIS-0, XIS-1, XIS-2, and XIS-3; \citealt{Koyama2007}) as focal plane
imager of four X-ray telescopes, and a non-imaging instrument called
the Hard X-ray Detector (HXD; \citealt{Takahashi2007}) consisting of
Si PIN photo-diodes and GSO scintillation counters. XIS-0, XIS-2, and
XIS-3 are front-side illuminated CCDs (FI-XISs), while XIS-1 is the
back-side illuminated one (BI-XIS).

In this paper, we analyze the data of the XISs and the HXD/PIN
(HXD nominal position), which
covers the energy band of 0.2--12 keV and 10--60 keV, respectively,
since the fluxes of our targets above 50 keV are too faint to be
detected with the HXD/GSO. Table \ref{tab-observations} shows the log
of the observations. The net exposure of \swtwo\ and \swfive\ is about
100 ks and 80 ks, respectively, while that for the rest 4 targets is
20 ks each\footnote{The primary goal of the first two targets is to
  constrain the reflection component with the best accuracy, aiming at
  the brightest new sources in the (then) latest BAT catalog. For the
  four targets with 20 ks exposure, the main goal was to obtain their
  X-ray spectra below 10 keV to make the BAT sample
  complete.}. 
Because XIS-2 became unoperatable on 2007 November 7
\citep{Dotani2007}, no XIS-2 data are available for \swtwo\ and
\swthree .
For the observations of \swtwo\ and \swthree,
we applied spaced-row charge injection (SCI) for the XIS data
to improve the energy resolution \citep{Nakajima2008}.
In the spectral analysis, we also
utilize the BAT spectra covering the 15--200 keV band, integrated for
the first 9 months for \swone\ and \swsix, and for 22 months for the
rest of targets.

\section{Analysis and Results}

We analyze the \suzaku\ data using {\it HEAsoft} version 6.3.2 from
the data products version 2.0 distributed by the {\Suzaku} pipeline
processing team. In extraction of the light curves and spectra, we set
the source region as a circle around the detected position with a
radius of 1.5--2 arcmin, depending on the flux. For the XIS data, the
background was taken from a source-free region in the field of view
with an approximately same offset angle from the optical axis as the
source. For the HXD/PIN data, we use the so-called ``tuned''
background model provided by the HXD team. Its systematic errors are
estimated to be $\simeq 1.3\%$ at a $1 \sigma$ confidence level in the
15--40 keV band for a 10 ks exposure \citep{Mizuno2008}. As our
exposures are $\approx$20 ks or larger, we expect that the error is
even smaller than this value.

\subsection{Light Curves}

Figures~\ref{fig-lc_xis} and \ref{fig-lc_pin} show the
background-subtract light curves of our targets obtained with the XIS
and HXD in the 2--10 keV and 15--40 keV band, respectively. To
minimize any systematic uncertainties caused by the orbital change of
the satellite, we merge data taken during one orbit (96 minutes) into
one bin. Then, to check if there is any significant time variability
during the observations, we perform a simple $\chi^2$ test to each
light curve assuming a null hypothesis of a constant flux. The
resultant reduced $\chi^2$ value and the degree of freedom are shown
in each panel. We detect no significant time variability on a time
scale of hours for all the targets in both energy bands.
Thus, we analyze the spectra of all the targets averaged over the whole
observation epoch.

\subsection{BAT Spectra} \label{sec-BAT_spectra}

Before performing the spectral fit to the \suzaku\ data, we firstly
analyze only the BAT spectra in the 15--200 keV band to constrain the
high energy cutoff in the continuum. It is known that the incident
photon spectra of Seyfert galaxies are well approximated by a power
law with an exponential cutoff (cutoff power law model), represented
as $A E^{-\Gamma} \exp \{ -E/E_{\rm cut}\}$, where $A$, $\Gamma$,
$E_{\rm cut}$ is the normalization at 1 keV, photon index, cutoff
energy, respectively. Here we utilize the \texttt{\bf pexrav} code by
\citet{Magdziarz1995} to take into account possible contribution of
the Compton reflection component from optically thick, cold matter.
The relative intensity of the reflection component to that of the
incident cutoff power law component is defined as $R \equiv \Omega / 2
\pi$, where $\Omega$ is the solid angle of the reflector ($R = 1$
corresponds to the reflection from a semi-infinite plane).

In this stage, we assume $R$ = 0 or 2 as the two extreme cases just to
evaluate the effects of the reflection.  The inclination angle is
fixed at $60^{\circ}$.  To avoid strong coupling between the power law
slope and and cutoff energy, we fix the photon index at 1.9, which is
a canonical value for AGNs. Table~\ref{tab-ecut} summarizes the
fitting results of the cutoff energy for each target. While it is
difficult to constrain its upper limit due to the limited band
coverage of the BAT, we find the cutoff energy must be above
$\approx$100 keV in most cases. Considering this limitation, we fix
$E_{\rm cut}$ at 300 keV in all subsequent analysis.
Impacts by adopting a lower value of $E_{\rm{cut}}$ 
in the spectral fitting are discussed in~\S\ref{sec-results-summary}.

\subsection{Spectral Models}

We consider three basic models in the spectral analysis uniformly for
all the six targets. Our policy is to start with the simplest model
for each target; if we find that the fit with a simple model does not
give a physically self-consistent picture and/or that the fit is
significantly improved by introducing additional parameters, then we
adopt more complicated models. In all the cases, we assume a power law
with an exponential cutoff fixed at 300 keV for the incident
continuum, as explained in the previous subsection. The Galactic
absorption, $N_{\rm{H}}^{\rm{Gal}}$, is always included in the model
(even if not explicitly mentioned below) by assuming the hydrogen
column density from the H~I map of \citet{Kalberla2005}, available
with the {\it nh} program in the {\it HEAsoft} package. For
absorption, we adopt the photo-electric absorption cross section by
\citet{Balucinska1992} (``bcmc'') and use the \texttt{\bf phabs} or
\texttt{\bf zphabs} model in XSPEC. Solar abundances by
\citet{Anders1989} are assumed throughout the analysis.

The first (simplest) model, designated as Model~A, consists of (i) a
transmitted component (a cut-off power law absorbed by cold matter),
(ii) a scattered component (a cut-off power law without absorption),
and (iii) an iron-K emission line (a gaussian), represented as
\texttt{\bf zphabs*zhighect*zpowerlw + const*zhighect*zpowerlw +
zgauss } in the XSPEC terminology\footnote{the {\bf zhighect} model is
utilized to represent exponential cutoff at the source redshift.}.
According to the unified scheme, in type 2 AGNs, the incident power
law from the nucleus will be absorbed by a dusty torus around the
SMBH, while the nuclear emission is partially scattered into the
line-of-sight by ionized gas around the torus. For simplicity, we
assume that the scattered component has the same shape of the incident
power law with a fraction of $f_{\rm scat}$ as first-order
approximation, although, in reality, it often contains a number of
recombination lines produced by a photo-ionized plasma
\citep[e.g.,][]{Sako2000} unless the scatterer is fully ionized. As a
result of reprocessing by cold matter in the surrounding environment
(such as a torus, an ionized gas, and an accretion disk), an iron-K
emission will be produced at the rest-frame 6.4 keV. Since a line from
the torus should not resolved by the energy resolution of the XIS, we
fix the $1 \sigma$ line width of the iron-K line at the averaged value
of the (apparent) line width of the ${}^{55}\rm{Fe}$ calibration
source at 5.9 keV: 0.3 eV, 38 eV, 30 eV, 0.8 eV, 1.7 eV and 3.2 eV for
\swone, \two, \three, \five, \six, and \sixteen, respectively, to take
into account a possible systematic error in the energy response
\footnote{
The large line widths of \swtwo\ and \three\ are artifact caused by
inaccurate calibration for the data with SCI (\citealt{Matsumoto2007}; see 
{http://www.astro.isas.ac.jp/suzaku/analysis/xis/ver2.0/})
}.

In the second model, Model~B, we consider an additional contribution
of an absorbed Compton reflection component from optically thick
matter. This emission is expected from the inner wall of an (optically
thick) torus and/or the accretion disk, irradiated by the incident
continuum. Model~B is expressed as \texttt{\bf
zphabs*zhighect*zpowerlw + const*zhighect*zpowerlw + zgauss +
zphabs*pexrav } in the XSPEC terminology, where the last term
represents the reflection component (not including the direct one).
In the \texttt{\bf pexrav} model, we make the solid angle $\Omega$ of
the reflector seen from the nucleus as a free parameter, and fix the
inclination angle and the cutoff energy at $60^{\circ}$ and $300 \
\rm{keV}$, respectively. The photon index and normalization of the
power law are linked to those of the transmitted component. The
absorption to the reflection component is set to be independent of
that for the transmitted one, by considering a different geometry of
the emission region. To check the physical validity of Model~B, we
examine the equivalent width (E.W.) of the iron-K emission line with
respect to the reflection component, $\rm{E.W.}^{\rm{refl}}$, obtained
from the fit. Theoretically, it is expected to be 1--2 keV
\citep{Matt1991}. This value, however, tightly depends on the
geometries of the torus and the accretion disk. Thus, we regard the
result of Model~B as valid only if $\rm{E.W.}^{\rm{refl}} = 0.5-2 \
\rm{keV}$ but discard it otherwise.

We finally consider the third model (Model~C), where we assume two
different absorptions with different covering factors for the
transmitted component. This reduces the contribution of a less
absorbed reflection component compared with the case of a single
absorption as assumed in Model~B. In fact, simulations show that the
absorber in the torus can be patchy \citep{Wada2002}, resulting
in a time-averaged spectrum that is better modeled by multiple
absorptions to the transmitted continuum rather than by a single
absorber. In the XSPEC terminology, Model~C is expressed as
\texttt{\bf 
zphabs*zpcfabs*zhighect*zpowerlw + const*zhighect*zpowerlw + zgauss + zphabs*pexrav},
where the multiple terms of the first component
(absorbed, partial covering model) represents the two different
absorptions.

To summarize, we can write the three models of the photon spectrum $F \left( E \right)$ 
without the Galactic absorption as follows:
\begin{eqnarray}
 & F \left( E \right) = f_{\rm c} \exp \left\{ - N_{\rm H}^{\rm a} \sigma \left( E \right) \right\} I \left( E \right) + \left( 1 - f_{\rm c} \right)  \exp \left\{ - N_{\rm H}^{\rm b} \sigma \left( E \right) \right\} I \left( E \right) \nonumber \\
 & + f_{\rm{scat}} I \left( E \right) + G \left( E \right) + \exp \left\{ - N_{\rm{H}}^{\rm{refl}} \sigma \left( E \right) \right\} C \left( E \right) + S \left( E \right),
\end{eqnarray}
where 
$f_{\rm c}$ is the fraction of the more heavily absorbed component ($f_{\rm c} = 1$ in Models A and B),
$N_{\rm H}^{\rm a}$ and $N_{\rm H}^{\rm b}$ are the absorption column densities for the
transmitted component ($N_{\rm H}^{\rm b}=0$ in Models A and B), 
$\sigma \left( E \right)$ is the cross section of photoelectric absorption, 
$I \left( E \right) \equiv A E^{- \Gamma} \exp \{ -E/E_{\rm cut} \}
$ is the intrinsic cutoff power law component,
$f_{\rm{scat}}$ is the scattered fraction, 
$G \left( E \right)$ is a gaussian representing the iron-K emission line, 
$C \left( E \right)$ is the Compton reflection component ($C \left( E \right)=0$ in Model~A), 
and $S \left( E \right)$ represents additional soft components (see below).

\subsection{Fitting Results}

\begin{figure*}
\epsscale{0.8}
\plotone{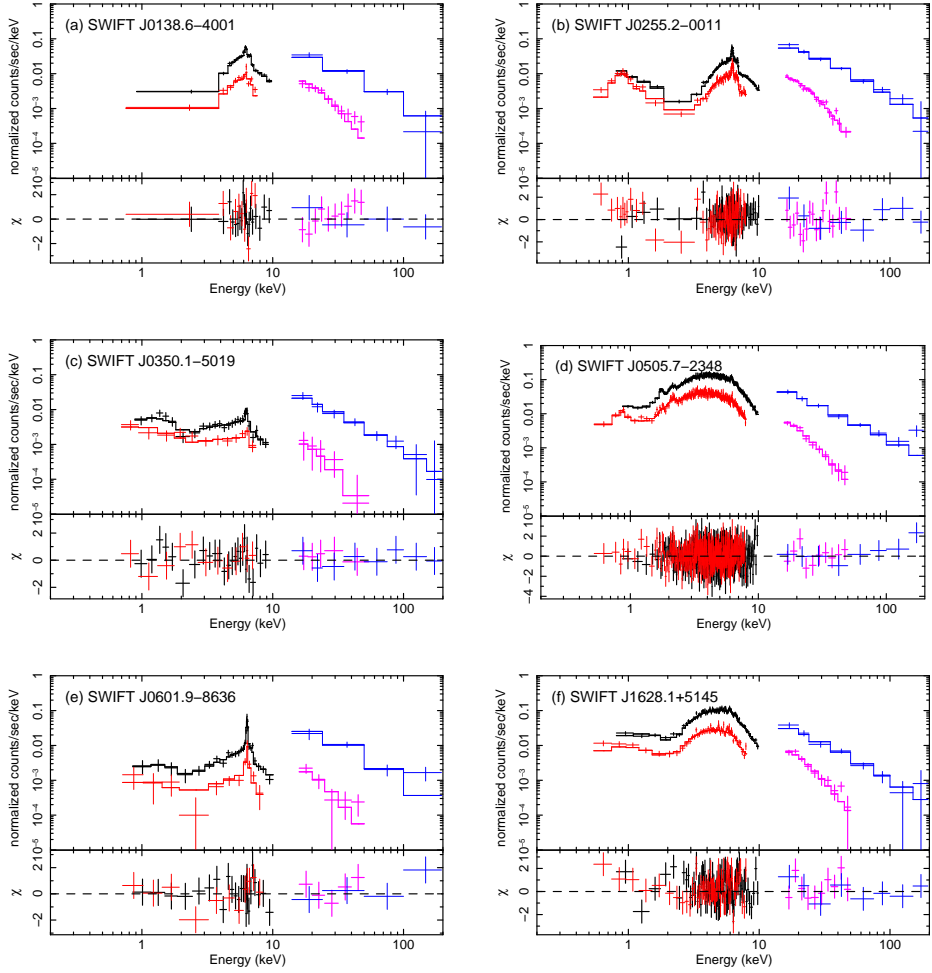}
\caption{
The observed spectra of the 6 targets.  The black, red, magenta and blue
crosses represent the data of the FI-XIS, BI-XIS, HXD/PIN, and BAT,
respectively, showing their 1$\sigma$ errors in vertical
direction. The spectra of the XIS and PIN are folded with the detector
response in units of counts s$^{-1}$ keV$^{-1}$, while those of the
BAT is corrected for the detector area and have units of photons
cm$^{-2}$ ks$^{-1}$ keV$^{-1}$. The best-fit models are plotted by lines, and the
residuals in units of $\chi$ are shown in the lower panels.
\label{fig-spectra_observed}}
\end{figure*}

Using the three models described in the previous subsection, we
perform spectral fit simultaneously to the spectra of the FI-XISs
(those of 2 or 3 XISs are summed), the BI-XIS, and the HXD/PIN. Based
on the fitting results, we chose the best model in the following
manner. (1) First, if Models~B is found to significantly improve the
fit from Model~A by performing an F-test, we tentatively adopt Model~B
as a better model, otherwise Model~A. (2) Then, if Model~B is found to
be physically {\it not} self-consistent in terms of the E.W. of the
iron-K line as explained in the previous subsection, or if Model~C
significantly improves the fit compared with Model A, we adopt Model~C
as the best model.

After finding the best one among the three models, we finally include
the \swift /BAT spectrum to the \suzaku\ data to constrain the photon
index most tightly, assuming that the spectral variability is
negligible between the observation epoch of \suzaku\ and that of
\swift. In the fit, we fix the relative normalization between the
FI-XISs and the PIN at 1.1 based on the calibration of Crab Nebula
\citep{Ishida2007}, while those of the BI-XIS and the BAT against
FI-XISs are treated as free parameters, considering the calibration
uncertainty and time variability (in flux), respectively. The detailed
results of spectral fit for each source is summarized below.

\subsubsection*{\swone}

We adopt Model B as the most appropriate model of \swone. We obtain
$(\chi^{2}, \nu)$ = $(101.14,\ 96)$ with Model A and $(92.52,\ 94)$
with Model~B, where $\nu$ is the degree of freedom. Thus, the
improvement of the fit by adding a reflection component is found to be
significant at $>$90\% confidence level by an $F$-test, which gives an
$F$-value of 4.38 for the degrees of freedom of $(2,\ 94)$. Since the
E.W. of the iron-K line with respect to the reflection component is
found to be $\rm{E.W.}^{\rm{refl}} >$ 0.5 keV, the model is
physically self-consistent. Here we allow the absorption to the
reflection component $N_{\rm{H}}^{\rm{refl}}$ to be a free parameter, unlike
the case in \citet{Ueda2007}, where it is linked to that for the
transmitted component. The basic results are essentially unchanged,
however.

\begin{figure*}
\epsscale{0.8}
\plotone{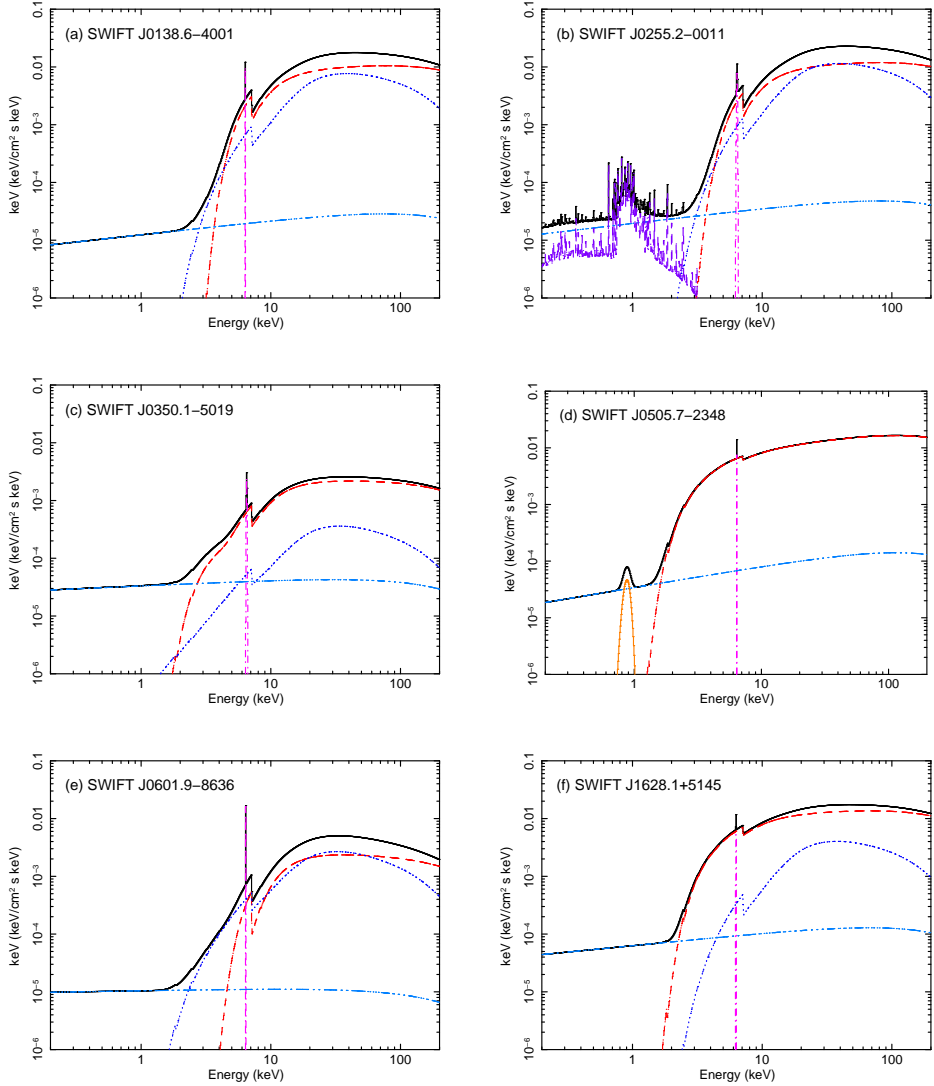}
\caption{
The best fit spectral model of the 6 targets in units of $E F_E$
(where $E$ is the energy and $F_E$ is the photon spectrum). The black,
dashed red, dotted blue, dot-dot-dashed cyan, dot-dashed magenta curves
correspond to the total, transmitted one, reflection component
, scattered component, and iron-K emission line, respectively.
The purple dashed model in \swtwo\ represents the
emission from an optically-thin thermal plasma, and the orange one in
\swfive\ the Ne emission lines from a photo-ionized plasma (see text).
\label{fig-spectra_model}}
\end{figure*}

\subsubsection*{\swtwo}

We adopt Model B with a {\it apec} plasma model for \swtwo.  Models~A
and B yield $(\chi^{2},\ \nu)$ = $(596.85,\ 304)$ and $(500.75,\
302)$, respectively, suggesting the presence of a significant amount
of a reflection component. We find $\rm{E.W.}^{\rm{refl}}
=0.81^{+0.07}_{-0.09}$ keV with Model B, which is physically
consistent. We also confirm that applying Model~C does not improve the
fit significantly, yielding $f_c = 1.0$ (i.e., multiple absorbers for
the transmitted component are not required when we consider a
reflection component).  As positive residuals remain in the energy
band below 1 keV, we add the {\it apec} model\footnote{
{http://cxc.harvard.edu/atomdb/}} in XSPEC, an emission model from an
optically-thin thermal plasma with Solar abundances, over Model B. It
greatly improve the fit, giving $(\chi^{2},\nu) = (330.38,\ 300)$. The
temperature of the plasma is found to be $kT = 0.74^{+0.03}_{-0.10} \
\rm{keV}$ with an emission measure of $n^2 V \approx 5 \times 10^{63}
\ \rm{cm}^{-3}$.  This temperature is similar to those found in many
Seyfert 2 galaxies \citep{Turner1997}.

\subsubsection*{\swthree}

Model C is adopted for {\swthree} since we find $\rm{E.W.}^{\rm{refl}}$
$< 0.1 \ \rm{keV}$ . We obtain $(\chi^{2}, \nu)$ = $(34.93,\ 36)$ with
Model A, and $(28.24,\ 35)$ with Model B. Here, we link the absorption
to the reflection component and that to the transmitted one, as having
the former as a free parameter does not help improving the fit.
Model~C gives $(\chi^{2}, \nu)$ = $(28.24,\ 35)$ and
$\rm{E.W.}^{\rm{refl}} = 2.8^{+1.4}_{-1.3}$ keV, which is now
physically consistent.

\subsubsection*{\swfive}

We adopt Model~C with a line feature at $\approx$0.9 keV as the best
description for \swfive. We find very little reflection component in
the spectra of \swfive ; Models~A and B give a very similar $\chi^{2}$
value of 671.3 for a degree of freedom of 602 and 600, respectively,
with no improvement of the fit by Model B. In fact, we obtain a tight
upper limit for the reflection component of $R < 0.11$ (90\%
confidence level) from the fit with Model~B. Thus, we apply Model~C
(here $N_{\rm{H}}^{\rm{refl}}$ is tied to $N_{\rm{H}}^{\rm{b}}$ since
$R \simeq 0$). This model yields $(\chi^{2},\ \nu) = (636.11,\ 598)$,
which is significantly better than Model~A (or B).
Furthermore, positive residuals remain in the 0.7--1
keV band, centered at $\simeq 0.9 \ \rm{keV}$. To model this feature,
we firstly add an {\it apec} component to Model~C like in the case of
\swtwo. This is found to be unsuccessful, largely overestimating the
1--1.2 keV flux of the data. Rather, in some obscured AGNs, emission
lines from Ne ions in a photo-ionized plasma are observed around 0.9
keV (e.g., NGC~4507 in \citealt{Comastri1998} and NGC~4151 in \citealt{Ogle2000}),
which is a likely origin of the excess feature seen in our spectra. As
these emission lines cannot be resolved by CCDs, we model it with a
gaussian whose $1\sigma$ width is fixed at 0.05 keV, referring to the
result of the Seyfert 2 galaxy NGC~4507 \citep{Matt2004}. The center
energy of the gaussian is treated as a free parameter, and is found to
be $0.87\pm0.02 \rm{keV}$, consistent with our assumption. We obtain
$(\chi^{2},\ \nu) = (615.08,\ 597)$, which is a significantly better
fit compared with Model C without the line.

\subsubsection*{\swsix}

Model B is adopted for \swsix. We confirm the results reported in
\citet {Ueda2007}. Models A and B yield $(\chi^{2}, \nu)$ = $(45.20,\
37)$ and $(32.32,\ 35)$, respectively, suggesting a significant
amount of a reflection component. As $\rm{E.W.}^{\rm{refl}}$ is found
to be $2.10^{+0.28}_{-0.30}$ keV, we find Model B to be a physically
consistent model for this source.

\subsubsection*{\swsixteen}

Model C is adopted for \swsixteen.  Model A gives $(\chi^{2}, \nu)$ =
$(574.66,\ 546)$, which is significantly improved to $(564.23,\ 544)$
in Model B but with $\rm{E.W.}^{\rm{refl}}$ of $<$ 0.1 keV. Hence, we
apply Model~C to the data, which yields $(\chi^{2}, \nu)$ = $(553.88,\
543)$ and a physically consistent value of $\rm{E.W.}^{\rm{refl}}$ =
$0.87^{+0.34}_{-0.35}$ keV.

\subsection{Results Summary} \label{sec-results-summary}

Table~\ref{tab-parameters} summarizes the choice of the spectral model
(A, B, or C) and best-fit parameters from the \suzaku\ + BAT
simultaneous fit for each target. The observed fluxes in the 2--10 keV
and 10--50 keV bands and the estimated 2--10 keV intrinsic luminosity
corrected for absorption are also listed. We find that all the six
targets are heavily obscured with column densities larger than
$>3\times10^{23}$ cm$^{-2}$. As already reported by \citet{Ueda2007},
\swsix\ can be called a mildly ``Compton thick'' AGN, since it shows
log $N_{\rm H} > 24$ cm$^{-2}$ but a transmitted component is still
seen in the hard X-ray band above 10 keV. We confirm that the photon
indices are within the range of 1.6--2.0 for all the targets, and
hence the fitting model is physically proper.

The upper panels of Figure~\ref{fig-spectra_observed} show the
observed spectra of the FI-XIS (black), the BI-XIS (red), and the
HXD/PIN (magenta) folded with the detector response in units of counts
s$^{-1}$ keV$^{-1}$, together with the BAT spectra (blue) corrected
for the detector area in units of photons cm$^{-2}$ ks$^{-1}$ keV$^{-1}$.
The best-fit models are superposed by solid lines. The lower panels show
the residuals in units of $\chi$ (i.e., divided by the $1\sigma$
statistical error in each bin). Figure~\ref{fig-spectra_model} shows
the incident spectra without Galactic absorption in units of $E
F_{E}$, where the contribution of each component in equation (1) is
plotted separately; the black, red, blue, cyan, magenta curves
correspond to the total, transmitted component, reflection component,
scattered component, and iron-K emission line, respectively. For
\swtwo\ and \swfive , the additional soft components are also included
in purple and orange, respectively.

For three sources \swone, \swtwo, and \swsix, the best fit 
reflection strength parameter $R (\equiv \Omega/2\pi) \gtrsim 1$, although
values close to unity are allowed within the uncertainties.
Most simple geometries only allow $R \approx 1$
to be the maximum physically plausible value. The most likely
explanation for these possibly unphysical results is that a part of
the direct emission from the nucleus is completely blocked by
nonuniform material in the line of sight, reducing the observed
normalization of the transmitted component smaller than the true value
by a factor of $\lesssim 1/R$ \citep{Ueda2007}. Alternatively, time
variability can be responsible for this, if we are observing an echo
of previously brighter phases of the AGN in the reflection
component. Thus, we list in Table~\ref{tab-parameters} the intrinsic
luminosities corrected for either of these effects by multiplying
$R$. Accordingly, the scattered fraction $f_{\rm{scat}}$ is corrected
by $1/R$ from the observed value, even though this correction does not
affect our conclusion.

To evaluate {\it maximum} possible errors in the spectral parameters
due to the uncertainty of the cutoff energy, we also perform fitting
by adopting $E_{\rm{cut}}$ = 100 keV instead of 300 keV for \swthree,
\swsix, and \swsixteen, whose 90\% lower limits of $E_{\rm{cut}}$ are
smaller than 100 keV as determined with the \swift /BAT spectra
(\S~\ref{sec-BAT_spectra}). We find that \swthree\ shows a smaller
scattered fraction ($f_{\rm scat}=0.77^{+0.58}_{-0.15}$\%) compared
with the case of $E_{\rm cut}$= 300 keV. For \swsixteen, we also
obtain a slightly smaller scattered fraction ($f_{\rm
scat}=0.74\pm0.06$) with stronger reflection
($R=0.80^{+0.17}_{-0.23}$) and larger absorption ($N_{\rm
H}^a=(87\pm14)\times10^{22} {\rm cm}^{-2}$). The other parameters,
including those of \swsix, do not change within the 90\% errors. These
results indicate that coupling of the spectral parameters with $E_{\rm
cut}$ could not always be negligible. Nevertheless, we confirm that it
does not affect our discussion below, even considering the largest
uncertainties as estimated here.

\section{Discussion}

\begin{figure}[b]
\epsscale{1.0}
\plotone{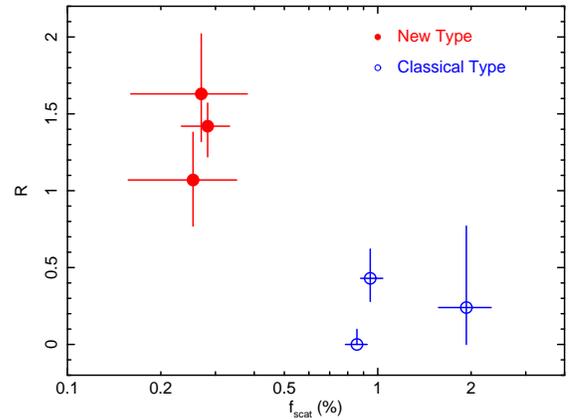}
\caption{
The correlation between the strength of the Compton reflection
component ($R=\Omega/2\pi$) and the fraction of the scattered
component ($f_{\rm{scat}}$) for the 6 targets.  Filled and open
circles represent ``new type'' ($R \gtrsim 0.8$ and $f_{\rm{scat}}
\lesssim 0.5\%$) and ``classical type'' ($R \lesssim 0.8$ and
$f_{\rm{scat}} \gtrsim 0.5\%$), respectively.
\label{fig-R_vs_fscat}}
\end{figure}

We have for the first time obtained broad-band spectra covering the
0.5--50 keV band of the six ``new'' AGNs detected in the \swift /BAT
survey that did not have precise X-ray observations, including the two
sources already reported by \citet{Ueda2007}. These six targets were
essentially selected without biases, and hence can be regarded as a
representative sample of new \swift /BAT AGNs.  The spectra of all the
targets are uniformly described with a spectral model consisting of a
heavily absorbed (log $N_{\rm H} > 23.5$ cm$^{-2}$) transmitted
components with a single or multiple absorptions, scattered lights, a
Compton reflection component from optically-thick cold matter, and an
iron-K emission line at 6.4 keV in the rest-frame, with additional
soft X-ray components in two cases. Thanks to its good sensitivity in
the 10--50 keV band and simultaneous band coverage with \suzaku, we
have accurately derived key spectral parameters including the fraction
of the scattered component $f_{\rm scat}$, the strength of the
reflection component $R(=\Omega/2\pi)$, and the E.W. of the iron-K
emission line.

The first notable result is that our sample show very small values of
$f_{\rm{scat}}$, $< 3\%$, compared with a typical value of optical
selected Seyfert-2 samples of 3--10\%
\citep[e.g.,][]{Turner1997,Cappi2006,Guainazzi2005}. \citet{Winter2008b}
categorize AGNs with $f_{\rm scat} < 3\%$ as ``hidden'' AGNs, to which
all the six sources belong. This is not unexpected, since the
``hidden'' AGNs constitutes a significant fraction (24\%) of the
uniform \swift /BAT sample.

Figure~\ref{fig-R_vs_fscat} shows the correlation between the
reflection strength $R$ versus scattering fraction $f_{\rm{scat}}$ for
our sample. This plot suggests that there are two groups even
within the ``hidden'' population defined by \citealt{Winter2008b},
that is,
\begin{itemize}
\item $R \gtrsim 0.8$ and $f_{\rm{scat}} \lesssim 0.5\%$
\item $R \lesssim 0.8$ and $f_{\rm{scat}} \gtrsim 0.5\%$,
\end{itemize}
although it is hard to conclude if this distribution is 
really distinct or continuous, given the small number of the current sample.
Following the description by \citet{Ueda2007}, we refer to the
former group as ``new type'' and the other ``classical type''. The
new group includes the two sources of \citet{Ueda2007},
\swone\ and \swsix, and \swtwo. The extremely small
scattering fraction indicates two possibilities (1) the opening angle
of the torus is small and/or (2) there is very little gas around 
the torus to scatter the X-rays.
Although it is difficult to firmly exclude
the latter possibility, the global presence of a reflection component
strongly supports the former, as discussed below.

We can constrain the geometry of the torus and viewing angles from the
spectral parameters, the scattering fraction, the E.W. of the iron-K
line, and absorption column density, based on a simple torus
model. Here we utilize the calculation by \citet{Levenson2002}, where
three free parameters are introduced for a uniform-density torus whose
cross-section has a square shape; the thickness of the torus in terms
of the optical depth for Thomson scattering $\tau$, the half-opening
angle of torus $\theta$, and the inclination angle $i$.  In this
geometry, the line-of-sight column density of the torus depends on
$i$; when we see the nucleus with face-on view, the absorption becomes
much smaller than the case of edge-on view. 
If the column density of the scatterer is constant, the opening 
angle can be connected to the scattering fraction as 
\begin{equation}
\cos \theta = 1 - \frac{f_{\rm{scat}}}{f_{\rm{scat,0}}} \left( 1 - \cos \theta_{0} \right). \label{eq-theta}
\end{equation}
where we normalize $f_{\rm{scat}}=f_{\rm{scat},0}$ at $\theta =
\theta_{0}$. Assuming $f_{\rm{scat,0}} = 3\%$ at $\theta_{0} = 45^{\circ}$
as typical parameters
\footnote{The value $\theta_{0} = 45^{\circ}$ corresponds to the
number ratio of obscured AGNs to unobscured ones of about 3, which is
consistent with the observations of local AGNs.},
we obtain $\theta \sim 20^{\circ}$ for the new type group. Then, 
comparing with Figure~2 of \citet{Levenson2002} where $\tau=4$ is assumed,
we can constrain $i \sim 40^{\circ}$ for
\swsix, and $i \sim 20^{\circ}$ for \swone\ and \swtwo, 
from the observed E.W. of the iron-K line
(here we have neglected the contribution of an iron-K line originating
from the accretion disk, which is often broader than the
``narrow'' one considered in the spectral model.). The absorption
column densities $N_{\rm{H}}$ $\gtrsim 10^{23.8} \ \rm{cm}^{-2}$ and
strong reflection $R\gtrsim 1$ are fully consistent with the geometry
considered here. The reflection continuum is attributable to that from
the inner wall of the tall torus and that from the accretion disk seen
with a small inclination angle.

Thus, it is very likely that we are seeing the new type AGNs in a
rather face-on geometry through a geometrically and optically thick
torus, as discussed in \citet{Ueda2007}. An important consequence of
this interpretation is that, if we view the same objects in a more
edge-on geometry, then these AGNs look heavily Compton-thick ($N_{\rm
H} \sim 10^{25}$ cm$^{-2}$) and hence the transmitted components
hardly escape toward us even in the hard X-ray $>10$ keV regime due to
repeated scattering. Thus, simliar systems to new type AGNs discovered
here may still be largely missed even in the on-going {\swift}/BAT or
\integral\ surveys, requiring even more sensitive observations in hard
X-rays to detect them more completely.

For the classical type of AGNs, we obtain $\theta \gtrsim 30^{\circ}$
from the observed scattered fraction by equation (\ref{eq-theta}).
Two sources in this group, \swfive\ and \swsixteen, show weak E.W. of
the iron-K emission line, $\sim 0.05$ keV.
This is hard to explain by a torus with large $\tau$ as
assumed in Figure~2 of \citet{Levenson2002}, which predicts
E.W. $>0.1$ keV if $i > \theta$ (i.e., the transmitted component is
absorbed). To consider a torus with different optical depths, we apply
a model developed by \citet{Ghisellini1994} (see their Figure~1
for the definition of the geometry). We find that the E.W of the
iron-K line and absorption can be consistently explained, within a
factor of 2, if the column density of the torus in the equatorial
plane is $\sim 10^{23-24}$ cm$^{-2}$ and $60^{\circ} \le i \le 84^{\circ}$. 
The absence of strong reflection
components is also consistent with the small optical depth of the
torus and with the ``edge-on'' accretion disk.

\begin{figure}[b]
\epsscale{1.0}
\plotone{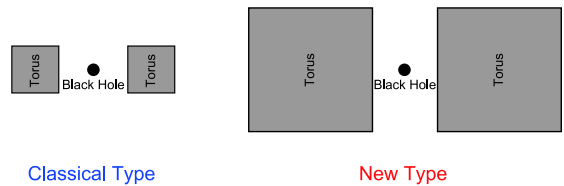}
\caption{
The schematic illustration of the torus geometry for the classical
type AGNs (left panel) and the new type AGNs (right panel).
\label{fig-torus}}
\end{figure}

To summarize, we have discovered from our first \swift-\suzaku\ AGN
sample two different classes of obscured AGN population, called ``new
type'' that have an extremely small scattering fraction ($f_{\rm scat}
< 0.5\%$) and a strong reflection component ($R \gtrsim 0.8$), and
``classical type'' with $f_{\rm scat} > 0.5\%$ and $R \lesssim
0.8$. Figure~\ref{fig-torus} shows a schematic illustration of the
torus geometry of the two types. It is likely that the new type AGNs
are deeply buried in an optically thick torus ($N_{\rm H} \sim
10^{25}$ cm$^{-2}$) with a small opening angle $\theta \sim 20^\circ$,
and are viewed in a face-on geometry. By contrast, the classical type
AGNs have an optically thin torus ($N_{\rm H} \sim 10^{23-24}$
cm$^{-2}$) with a larger opening angle $\theta \gtrsim 30^\circ$ viewed
in a more edge-on geometry than the new type. A significant number of
new type AGNs with an edge-on view from us may still be missing in the
current all-sky hard X-ray surveys. The presence of two classes of
obscured AGNs implies that the understanding of AGN is not as simple
as assumed in the population synthesis models of the cosmic X-ray
background. Given the small number of our sample, however, it is not
clear whether there are two ``distinct'' types of obscured AGNs or the
distribution is continuous. More systematic follow-up observations
with \suzaku\ of new {\swift}/BAT AGNs are important to establish the
true properties of local AGNs.

\acknowledgments

We thank Lisa Winter for useful comments to the manuscript. Part of
this work was financially supported by Grants-in-Aid for JSPS Fellows
(SE), for Scientific Research 20540230 and 20740109, and for the
Global COE Program ``The Next Generation of Physics, Spun from
Universality and Emergence'' from from the Ministry of Education,
Culture, Sports, Science and Technology (MEXT) of Japan. This research
has made use of the NASA/IPAC Extragalactic Database (NED) which is
operated by the Jet Propulsion Laboratory, California Institute of
Technology, under contract with the National Aeronautics and Space
Administration.

\clearpage
\LongTables
\begin{landscape}
\begin{deluxetable}{cccccccc}
\tablenum{4}
\tablecaption{Best-fit Spectral Parameters\label{tab-parameters}}
\tablewidth{0pt}
\tablehead{\colhead{} & \colhead{SWIFT} & \colhead{J0138.6--4001} & \colhead{J0255.2--0011} & \colhead{J0350.1--5019}
 & \colhead{J0505.7--2348} & \colhead{J0601.9--8636} & \colhead{J1628.1$+$5145}}\startdata
 & Best-fit model & B & B + apec\tablenotemark{a} & C & C\tablenotemark{b} & B & C \\
(1) & $N_{\rm{H}}^{\rm{Gal}}$ ($10^{22} \ \rm{cm}^{-2}$) & $0.0163$ & $0.0581$ & $0.0116$ & $0.0212$ & $0.102$ & $0.0183$ \\
(2) & $N_{\rm{H}}$ or $N_{\rm{H}}^{\rm{a}}$ ($10^{22} \ \rm{cm}^{-2}$) & $65.1^{+4.2}_{-4.6}$ & $63.1^{+2.1}_{-1.8}$ & $85 \pm 16$ & $29.3 \pm 6.2$ & $115^{+23}_{-18}$ & $58 \pm 11$ \\
(3) & $N_{\rm{H}}^{\rm{b}}$ ($10^{22} \ \rm{cm}^{-2}$) & --- & --- & $8.8^{+3.0}_{-2.1}$ & $5.868 \pm 0.081$ & --- & $12.91^{+0.36}_{-0.45}$ \\
(4) & $f_{\rm{c}}$ & --- & --- & $0.877^{+0.033}_{-0.037}$ & $0.261^{+0.057}_{-0.023}$ & --- & $0.380^{+0.036}_{-0.039}$ \\
(5) & $\Gamma$ & $1.755^{+0.081}_{-0.048}$ & $1.7778 \pm 0.022$ & $1.900^{+0.064}_{-0.070}$ & $1.621^{+0.021}_{-0.032}$ & $1.962^{+0.066}_{-0.064}$ & $1.799^{+0.030}_{-0.029}$ \\
(6) & $f_{\rm{scat}}$ (\%) & $0.276^{+0.070}_{-0.068}$ & $0.263 \pm 0.046$ & $2.03^{+0.38}_{-0.41}$ & $0.857^{+0.067}_{-0.070}$ & $0.27 \pm 0.11$ & $0.946^{+0.074}_{-0.078}$ \\
(7) & $E_{\rm{cen}}$ (keV) & $6.375^{+0.026}_{-0.030}$ & $6.3911 \pm 0.0089$ & $6.40^{+0.20}_{-0.04}$ & $6.392 \pm 0.019$ & $6.401^{+0.013}_{-0.012}$ & $6.275^{+0.067}_{-0.051}$ \\
(8) & E.W. (keV) & $0.167 \pm 0.042$ & $0.218 \pm 0.020$ & $0.28^{+0.13}_{-0.14}$ & $0.062^{+0.012}_{-0.011}$ & $1.14 \pm 0.16$ & $0.051 \pm 0.021$ \\
(9) & $N_{\rm{H}}^{\rm{refl}}$ ($10^{22} \ \rm{cm}^{-2}$) & $8.5^{+2.8}_{-2.7}$ & $12.4^{+1.9}_{-1.2}$ & $7.3$ ($> 0$) & ($= N_{\rm{H}}^{\rm{b}}$) & $3.7^{+3.1}_{-1.4}$ & ($= N_{\rm{H}}^{\rm{b}}$) \\
(10) & $R$ & $1.00^{+0.36}_{-0.26}$ & $1.49^{+0.15}_{-0.23}$ & $0.28$ ($< 0.92$) & $1.4 \times 10^{-7}$ ($< 9.8 \times 10^{-2}$) & $1.63^{+0.39}_{-0.32}$ & $0.43^{+0.22}_{-0.16}$ \\
(11) & $F_{\rm{2-10}}$ ($\rm{ergs} \, \rm{cm}^{-2} \, \rm{s}^{-1}$) & $3.4 \times 10^{-12}$ & $4.2 \times 10^{-12}$ & $9.2 \times 10^{-13}$ & $1.1 \times 10^{-11}$ & $1.0 \times 10^{-12}$ & $9.0 \times 10^{-12}$ \\
(12) & $F_{\rm{10-50}}$ ($\rm{ergs} \, \rm{cm}^{-2} \, \rm{s}^{-1}$) & $3.4 \times 10^{-11}$ & $4.3 \times 10^{-11}$ & $5.5 \times 10^{-12}$ & $2.9 \times 10^{-11}$ & $1.1 \times 10^{-11}$ & $3.6 \times 10^{-11}$ \\
(13) & $L_{\rm{2-10}}$ ($\rm{ergs} \, \rm{s}^{-1}$) & $2.4 \times 10^{43}$ & $5.1 \times 10^{43}$ & $1.4 \times 10^{43}$ & $1.5 \times 10^{44}$ & $8.9 \times 10^{41}$ & $1.5 \times 10^{44}$ \\
 & $\chi^{2} / \rm{d.o.f.}$ & $94.42 / 97$ & $335.97 / 307$ & $27.28 / 39$ & $622.38 / 604$ & $35.48 / 38$ & $558.76 / 550$ \\
\enddata
\tablenotetext{a}{An additional emission from an optically-thin
thermal plasma with Solar abundances is required, modelled by the \texttt{\bf apec} code
with a temperature of $kT = 0.74^{+0.03}_{-0.10} \ \rm{keV}$ and an
emission measure of $5 \times 10^{63} \ \rm{cm}^{-3}$ (see text).}
\tablenotetext{b}{An emission line feature at $\simeq 0.9 \ \rm{keV}$ is required, 
probably that from Ne ions from a photo-ionized plasma (see text).}
\tablecomments{
(1) The hydrogen column density of Galactic absorption by \citet{Kalberla2005}.
(2) The line-of-sight hydrogen column density for the transmitted component.
For the double covering model (Model C), 
that for more heavily absorbed component is given.
(3) The smaller line-of-sight hydrogen column density for the transmitted component in Model C.
(4) The normalization fraction of the more absorbed component to the total transmitted one in Model C.
(5) The power-law photon index.
(6) The fraction of the scattered component relative to the intrinsic power law,corrected for the transmission efficiency of $1 / R$ if $R > 1$ (see text).
(7) The center energy of the iron-K emission line at the rest frame of the source redshift.
(8) The observed equivalent width of the iron-K line with respect to the whole continuum.
(9) The line-of-sight hydrogen column density for the reflection component.
(10) The relative strength of the reflection component to the transmitted one,
defined as $R \equiv \Omega / 2 \pi$, where $\Omega$ is the solid angle of the reflector viewed from the nucleus.
(11) The observed flux in the 2--10 keV band.
(12) The observed flux in the 10--50 keV band.
(13) The 2--10 keV intrinsic luminosity corrected for the absorption and the transmission efficiency of $1 / R$ if $R > 1$.
The errors are 90\% confidence limits for a single parameter.
}
\end{deluxetable}
\clearpage
\end{landscape}

\end{document}